\documentclass[twocolumn,prb,amsmath,amssymb,showpacs,superscriptaddress,floatfix]{revtex4}
\usepackage{graphicx}
\usepackage{color}
\begin{document}
\title{Interaction-tuned compressible-to-incompressible phase transitions in the quantum Hall systems}
\author{Z. Papi\'{c}}
\affiliation{Laboratoire Pierre Aigrain, Ecole Normale Sup\'erieure, CNRS, 24 rue Lhomond, F-75005 Paris, France}
\affiliation{Laboratoire de Physique des Solides, Univ.~Paris-Sud, CNRS, UMR 8502, F-91405 Orsay Cedex, France}
\affiliation{Institute of Physics, P.~O.~Box 68, 11 000 Belgrade, Serbia}
\author{N. Regnault}
\affiliation{Laboratoire Pierre Aigrain, Ecole Normale Sup\'erieure, CNRS, 24 rue Lhomond, F-75005 Paris, France}
\author{S. Das Sarma}
\affiliation{Condensed Matter Theory Center, Department of Physics, \\
University of Maryland, College Park, MD 20742, USA}
\date{\today}
\newcommand{\red}{\color{red}}

\begin{abstract} 
We analyze transitions between quantum Hall ground states at prominent filling factors $\nu$ in the spherical geometry by tuning the width parameter of the Zhang-Das Sarma interaction potential. We find that incompressible ground states evolve adiabatically under this tuning, whereas the compressible ones are driven through a first order phase transition. Overlap calculations show that the resulting phase is increasingly well described by appropriate analytic model wavefunctions (Laughlin, Moore-Read, Read-Rezayi). This scenario is shared by both odd ($\nu=1/3, 1/5, 3/5, 7/3, 11/5, 13/5$) and even denominator states ($\nu=1/2, 1/4, 5/2, 9/4$). In particular, the Fermi liquid-like state at $\nu=1/2$ gives way, at large enough value of the width parameter, to an incompressible state identified as the Moore-Read Pfaffian on the basis of its entanglement spectrum.
\end{abstract}


\maketitle \vskip2pc

We address in this work, via large-scale exact diagonalization (ED) calculations on finite spheres, the important and interesting question of how to tune various fractional quantum Hall (FQH) ground
states between ungapped compressible and gapped incompressible phases by continuously varying the effective electron-electron interaction.  Such numerical studies have been a standard theoretical tool in FQH physics since the beginning\cite{laughlin} because of the non-perturbative nature of the FQH ground states.  In the current work, which is complementary to the pseudopotential description of quantum phase transitions (QPT) in quantum Hall systems as pioneered by Morf\cite{morf} and Haldane,\cite{haldane} we report that a simple single-parameter parametrization of the effective interaction through the so-called Zhang-Das Sarma (ZDS) model\cite{zds} provides a flexible and powerful method of studying QPTs between compressible and incompressible phases at both even and odd denominator FQH states.  We will show that ZDS interaction possesses a rich structure that can drive the FQH system from parameter regions where it appears to be compressible (manifested by the ground state that breaks rotational invariance i.e. the value of angular momentum $L \neq 0$) towards the incompressible region where the ground state is rotationally invariant ($L=0$), along with the corresponding overlap with the trial states like Laughlin\cite{laughlin} or paired states (Moore-Read Pfaffian\cite{mr}, Read-Rezayi\cite{rr_parafermion} etc.) jumping to a value close to unity and an energy gap opening up in the excitation spectrum. 
In agreement with the experimental phenomenology, we find that the well-known (small) odd-denominator incompressible FQH states (e.g. 1/3, 1/5, 7/3, 11/5) are robust and usually do not manifest any interaction-tuned QPT whereas the more fragile, even denominator (e.g. 1/2, 1/4, 5/2, 9/4) FQH states typically  exhibit characteristic QPT from a compressible to an incompressible phase as the Coulomb interaction is softened by increasing the ZDS tuning parameter. 

Our calculations are performed in the spherical geometry introduced and described in detail by Haldane,\cite{haldane} here we make only a few essential comments. We consider spin to be fully polarized and use the ZDS model interaction which was originally proposed to study the finite thickness effect of the quasi two-dimensional layer,\cite{zds} but in our analysis, the thickness parameter $w$ (expressed in units of the rescaled magnetic length, $l_B$) enters simply as the tuning parameter for the Hamiltonian: 

\begin{equation}\label{zds_interaction}
V_{\rm ZDS}(r)=\frac{1}{\sqrt{r^2+w^2}}.
\end{equation} 

We emphasize that the ZDS interaction (\ref{zds_interaction}) appears to have the same qualitative pseudopotential decomposition as the realistic models (e.g. the Fang-Howard, infinite square well, etc.), as has recently been shown in details in Ref.~\onlinecite{pjds}. However, it was also observed in Ref.~\onlinecite{papic_onefourth} that realistic confinement models do not always reproduce the QPTs induced by the ZDS interaction,  suggesting there may be subtle quantitative differences between ZDS and alternative confinement models which are important in the vicinity of a QPT. In this paper we focus on the ZDS model in carrying out our ED studies since a single parameter enables us to study FQH QPTs in a compact manner. In order to establish the connection with the experiments, we should mention that $w$ in the ZDS model corresponds roughly to the root-mean-square fluctuation in the electron coordinate in the transverse direction.\cite{pjds}

With this choice of the interaction, we use the overlap between the exact, numerically diagonalized finite  system,  and a candidate analytical wavefunction (e.g. the Laughlin or the Moore-Read wavefunction) to determine the tentative quantum phase of the system, i.e. if the overlap is `large' (`small'), the system is supposed to be in the candidate state (or not). We calculate the overlap as a continuous function of the varying Hamiltonian which is being tuned by $w$.  All the model wave functions studied in this paper are Jack polynomials that have squeezable configurations\cite{jack} which can be efficiently generated and compared with the exact ground state.
Note that each FQH state on a finite sphere at the filling factor $\nu$ is characterized, beside the number of electrons $N$ and the number of flux quanta $N_\phi$, also by a topological invariant $\delta$ called shift, defined by $N_\phi=\nu^{-1}N+\delta.$ In the thermodynamic limit of an infinite plane, the shift plays no role, but for a finite \emph{sphere} it is a crucial aspect of the ED technique\cite{haldane} as it can lead to an ``aliasing"\cite{morf_aliasing} problem: at a fixed choice of $(N_\phi, N)$, more than one quantum Hall state (having different $\nu$, $\delta$ and, therefore, different physical properties) may be realized. To avoid such loss of uniqueness for finite sphere ED, we disregard the aliased states from our considerations. Notwithstanding the aliasing problem, the system sizes we analyze are the largest that can be presently handled in ED studies. 

\begin{figure}[htb]
\centering
\includegraphics[width=\textwidth, angle=270, scale=0.35]{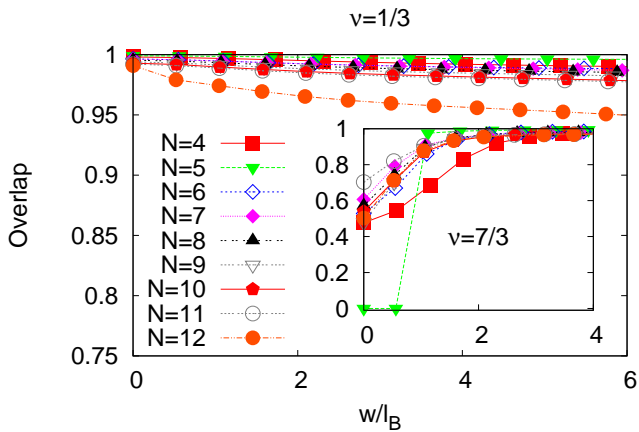}
\caption{(Color online) Overlap $|\langle \Psi_{\rm L}|\Psi_{\rm exact}\rangle|$ between the exact Coulomb state for finite width (ZDS model) at $\nu=1/3$ and the Laughlin wave function $N= 4 - 12$ particles. Inset: same quantity but in the first excited Landau level i.e. $\nu=7/3$.}
\label{fig_onethird}
\end{figure}

\begin{figure}[htb]
\centering
\includegraphics[width=\textwidth, angle=270, scale=0.35]{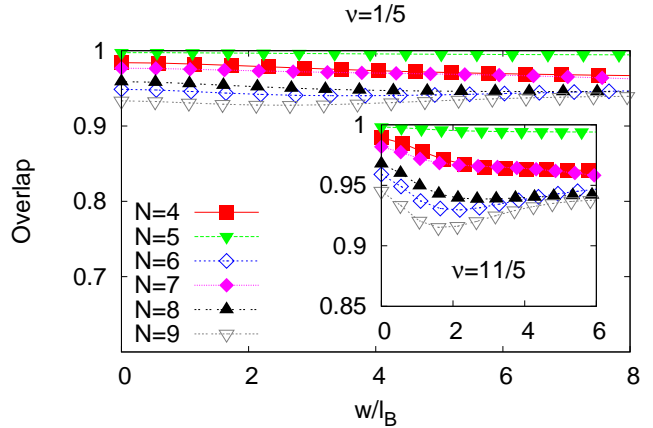}
\caption{(Color online) Overlap $|\langle \Psi_{\rm L}|\Psi_{\rm exact}\rangle|$ between the exact Coulomb state for finite width (ZDS model) at $\nu=1/5$ and the Laughlin wave function for $N=4 - 9$ particles. Inset: same quantity but in the first excited Landau level i.e. $\nu=11/5$.}
\label{fig_onefifth}
\end{figure}

We begin with the Laughlin fractions $\nu=1/3$ and $\nu=1/5$ in the lowest Landau level (LLL) and in the first excited Landau level ($\nu=2+1/3, 2+1/5$), Figs. \ref{fig_onethird}, \ref{fig_onefifth}. In agreement with previous studies,\cite{pjds} in the LLL we find that the ZDS potential leads to the monotonous decrease in the overlap with the Laughlin wave function with increasing the thickness parameter $w$.

In the second Landau level (SLL) and for zero thickness (Figs. \ref{fig_onethird}, \ref{fig_onefifth} inset) one first notices that the Laughlin $1/5$ wave function appears to be a better candidate than the one for $1/3$. Furthermore, certain particle numbers yield zero overlap for $\nu=7/3$ (for $N=5$ particles, the ground state is obtained in $L=2$ sector, therefore the overlap with the Laughlin wave function is zero due to the difference in symmetry). Things change once the ZDS potential is turned on: the states which are homogeneous ($L=0$) increase their overlap, while the finite-size artefact $N=5$ undergoes a quantum phase transition (QPT) turning into an $L=0$ state just under $w < l_B$.

\begin{figure}[htb]
\centering
\includegraphics[width=\textwidth, angle=270, scale=0.35]{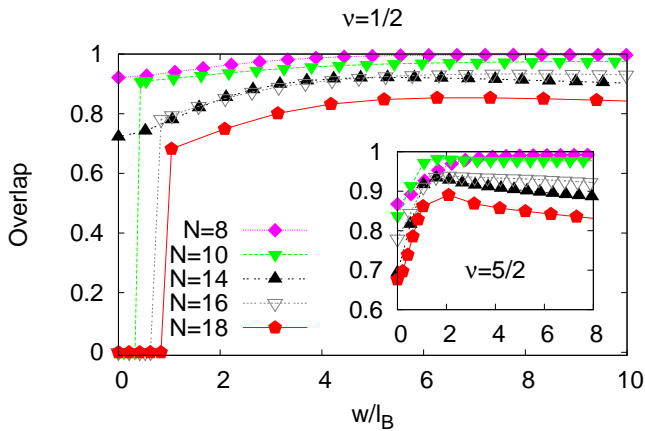}
\caption{(Color online) Overlap $|\langle \Psi_{\rm Pf}|\Psi_{\rm exact}\rangle|$ between the exact Coulomb state for finite width (ZDS model) at $\nu=1/2$ and the Pfaffian for $N=8 - 18$ particles.  Inset: same quantity but in the first excited Landau level i.e. $\nu=5/2$.  Only non-aliased states are shown. Note: the critical width of the QPT increases with system size, however for the three available points and $N\rightarrow \infty$, it extrapolates to a value of $4l_B$.}
\label{fig_onehalf}
\end{figure}

It will be shown in what follows that the induced QPT for $N=5, \nu=7/3$ is not an exceptional case.\cite{footnote} Even denominator fractions, such as $\nu=5/2$ which is believed to be the Moore-Read Pfaffian\cite{mr} or the recently discovered $\nu=1/4$,\cite{luhman} and various paired states of the Read-Rezayi sequence\cite{rr_parafermion} like $\nu=12/5, 13/5$, attract considerable attention because of their unusual ground states and the exotic spectrum of excitations that may be utilized in topological quantum computation.\cite{tqc} While their realization in the SLL seems a likely possibility, there has been little expectation to observe them in the conditions of the LLL (see however Ref~\onlinecite{shabani}.). In particular, at the thin single-layer $\nu=1/2$ in the LLL only the compressible, Fermi-liquid like state has been observed. In Fig.~\ref{fig_onehalf} we show the overlap results of finite-size calculations on $\nu=1/2$ in the LLL and $\nu=5/2$ in the SLL with ZDS interaction. 

At $\nu=1/2$ a QPT is induced by increasing the parameter $w$. Certain particle numbers yield good overlap already for zero thickness and their overlap will improve as $w$ increases. Other particle numbers produce ground states with well-defined values of $L>0$ that undergo a QPT at a critical value of the thickness. For $\nu=5/2$, the Coulomb ground state for zero thickness is already reasonably well approximated\cite{rh, ms} by the Moore-Read Pfaffian and the effect of ZDS interaction is only to increase the overlap in a smooth way. However, the increase is substantial -- up to $20\%$ for the largest system amenable to ED. This adiabatic continuity of the Moore-Read description for the SLL $\nu=5/2$ has been discussed in Ref.~\onlinecite{ms} and recently at great length by Storni \emph{ et al.}\cite{storni}

\begin{figure}[htb]
\centering
\includegraphics[scale=0.35, angle=270]{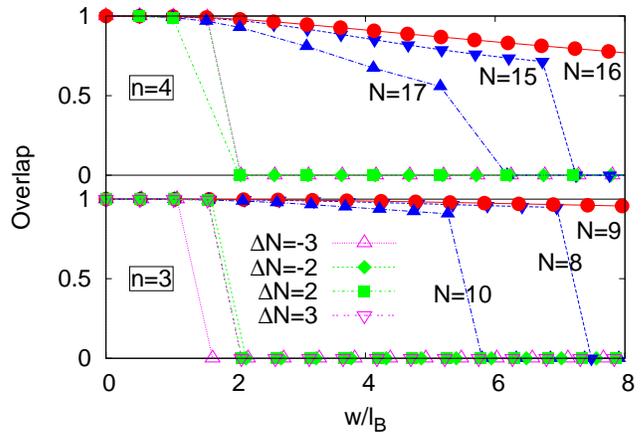}
\caption{(Color online) Overlap $|\langle \Psi_{\rm exact}(w=0)|\Psi_{\rm exact} (w) \rangle|$ between the exact Coulomb state for finite width (ZDS model) at $\nu=1/2$ and the CF sea state defined to be exact Coulomb ground state for zero thickness. Red circles represent filled CF shells ($N=n^2, n=3, 4$) , blue triangles are the lowest excited states $\Delta N=N-n^2=\pm 1$ and so on. }
\label{fig_cfsea}
\end{figure}

The non-zero values of $L$ that appear at $\nu=1/2$ in the LLL can be fully understood from the CF theory.\cite{morf_cfsea} Indeed, former work hinted at the possibility of $p-$wave paired CF state as a result of CF sea being perturbed by ZDS interaction.\cite{park} However, in Ref.~\onlinecite{park} only the variational energies of trial states were compared. In Fig. \ref{fig_cfsea} we will show that one can establish a connection between the ZDS-induced QPT and the Pfaffian and CF sea states in the LLL at $\nu=1/2$. 

Because the CF sea state and the Moore-Read Pfaffian occur at different shifts on the sphere (-2 and -3, respectively), one cannot simultaneously study their evolution with $w$. However, by analyzing the excitations of CF sea occuring at the Pfaffian shift, one can show (using Hund's rule) that the $L$ values obtained in ED \emph{at the Pfaffian shift} (Fig. \ref{fig_onehalf}) are indeed those stemming from the CF sea excitations.  Moreover, assuming that the Coulomb ground state in the LLL for zero thickness is exceedingly well approximated by Rezayi-Read wave function,\cite{rr} we define the CF sea state for our purposes as the interacting Coulomb ground state for zero thickness and study its overlap with the $w\geq 0$ ground states, Fig. \ref{fig_cfsea}. CF theory tells us that (at the shift of -2) the $L=0$ configurations are obtained when the CF shells are completely filled i.e. for $N=n^2, n=1,2,3,\cdots$ These configurations are particularly robust and adding/subtracting electrons from them ($\Delta N=N-n^2=\pm 1, \pm 2, \cdots$) creates a configuration that is destroyed at some critical value of the width which depends on how far away the system is from the filled shell. Obviously, there is ambiguity in defining precisely the critical width where the CF sea is destroyed, but this argument nonetheless provides further support for the claim that the ZDS-induced compressible--incompressible transition indeed proceeds via destruction of CF sea towards the Moore-Read Pfaffian. Transition of the same kind can be relevant for the multicomponent candidates\cite{scarola_rezayi_jain} at $\nu=3/8.$ We emphasize that the possible finite-width-induced LLL $\nu=1/2$ FQH state that we find arising out of the destabilization of the CF sea, even if it exists, is likely to be extremely fragile with a neutral excitation gap smaller than $0.03 e^2/\epsilon l_B$. \cite{storni} However, numerically extrapolated gap is generally known to be difficult to relate to the experimental value\cite{morf_ambrumenil} and in our data we cannot rule out the possibility that it goes to zero in thermodynamic limit.

\begin{figure}[htb]
\centering
\includegraphics[width=\textwidth, angle=270, scale=0.26]{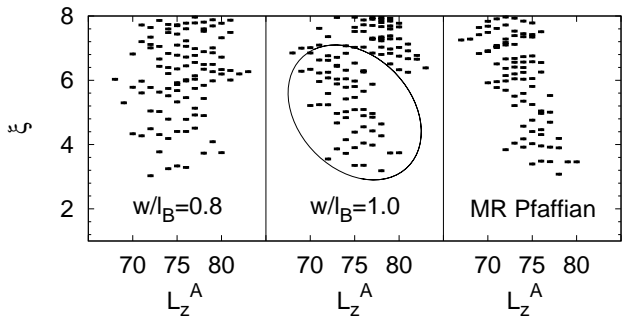}
\caption{Entanglement spectrum of the exact ground state for $N=18$ particles at $\nu=1/2$ in the LLL, just before ($w/l_B=0.8$) and after ($w/l_B=1.0$) the QPT, and the spectrum of Moore-Read Pfaffian for comparison. Vertical axes show the quantity $\xi=-\log  \lambda_A$, where $\lambda_A$ are the eigenvalues of the reduced density matrix of the subsystem $A$ which comprises of $8$ particles and $15$ orbitals, given as a function of angular momentum $L_z^A$. Data shown is only for the partitioning denoted by $\left[ 0|0 \right]$ in Ref.~\onlinecite{haldane_entspectrum}, other sectors give a similar result.}
\label{fig_entspectrum_16}
\end{figure}

Another way to look at the QPT towards the Moore-Read Pfaffian is to analyze the entanglement spectrum proposed in Ref.~\onlinecite{haldane_entspectrum}. This is a powerful way to identify topological order in the given ground state wave function and establish a direct connection with the underlying CFT that produces the given ground state as its correlator and thus offering more information than the simple overlap calculation.\cite{zhr} In Fig. \ref{fig_entspectrum_16} we show the change in the entanglement spectrum for $N=18$ particles at $\nu=1/2$ in the LLL, before and after QPT. For $w<l_B$, there is no visible CFT branch in the entanglement spectrum -- the generic Coulomb part dominates, leading to a likely compressible ground state. After the QPT, a CFT branch separates from the Coulomb part of the spectrum and the level counting begins to match the first few Virasoro levels of the Ising CFT. This is additional evidence in favor of the possibility of a finite-width-induced QPT to an incompressible half-filled single-layer LLL FQH state.

We have also examined the effect of ZDS potential on other even denominator and paired states. In the LLL, a QPT is induced for $\nu=1/4$\cite{papic_onefourth} around $w \sim 3 - 5 l_B$ and for Read-Rezayi $\nu=3/5$ state around $w \sim 4l_B$. In the SLL, a $\nu=9/4$ state is similarly stabilized when ZDS parameter is around $w \sim 3l_B$.  

Our work establishes that the continuous tuning of the interaction through the ZDS Hamiltonian enables a direct study of FQH quantum phase transitions showing that the usual odd denominator states are robust in both the LLL and the SLL, whereas the fragile even denominator FQH states are stable only in a regime of the interaction strength where the bare electron-electron interaction is considerably softer than the pure 2D Coulomb interaction. We find that the ZDS interaction allows for the existence of non-Abelian incompressible FQH states even at unusual even fractions such as $1/2, 1/4,$ and $9/4$, raising the intriguing possibility that such exotic non-Abelian states may indeed exist if one can sufficiently soften the interaction along the ZDS prescription.  Whether this can be physically achieved in 2D semiconductor systems remains an interesting open question and may require some `reverse engineering' of the quasi-2D samples to achieve a suitable density profile using the fact that the width parameter in the ZDS model corresponds roughly to the variance of the electron position in the transverse direction.

\section*{Acknowledgments}

This work was funded by the Agence Nationale de la Recherche under Grant
ANR-JCJC-0003-01. We also thank KITP for support through NSF Grant No. PHY05-51164. ZP was partly supported by the European Commission, through a Marie Curie Foundation contract MEST CT 2004-51-4307, and by the Serbian Ministry of Science under Grant No. 141035. SDS is supported by a Microsoft Q Research Grant.


\begin{thebibliography}{9}
\bibitem{laughlin}  R. B. Laughlin, Phys. Rev. Lett. {\bf 50}, 1395 (1983); Phys. Rev. B {\bf 27}, 3383 (1983).
\bibitem{morf} R. H. Morf, Phys. Rev. Lett. {\bf 80}, 1505 (1998).
\bibitem{haldane} F. D. M. Haldane, Phys. Rev. Lett. {\bf 51}, 605 (1983), see also \emph{The Quantum Hall Effect}, 2nd ed., edited by R. E. Prange and S. M. Girvin, Springer-Verlag, New York, 1990.
\bibitem{zds} F. C. Zhang and S. Das Sarma, Phys. Rev. B {\bf 33}, 2903 (1986).
\bibitem{mr} G. Moore and N. Read, Nucl.Phys. B {\bf 360}, 362 (1991).
\bibitem{rr_parafermion}  N. Read and E. Rezayi, Phys. Rev. B {\bf 59}, 8084 (1999).
\bibitem{pjds} M. Peterson, Th. Jolicoeur and S. Das Sarma, Phys. Rev. B {\bf 78}, 155308 (2008); Phys. Rev. Lett. {\bf 101}, 016807 (2008).
\bibitem{papic_onefourth} Z. Papi\'c, G. M\"oller, M. V. Milovanovi\'c, N. Regnault and M. O. Goerbig, Phys. Rev. B {\bf 79}, 245325 (2009).
\bibitem{jack} B. Andrei Bernevig and F. D. M. Haldane, Phys. Rev. Lett. {\bf 100}, 246802 (2008).
\bibitem{morf_aliasing} N. d'Ambrumenil and R. Morf, Phys. Rev. B {\bf 40}, 6108 (1989).
\bibitem{footnote} Another example where ZDS potential ``rectifies" the problem of compressibility is the supposedly interacting composite fermion (CF) state at $\nu=4/11$ where the ground state for $N=8$ electrons is compressible with $L=2$. We have verified that turning on ZDS potential in this case causes a QPT to the $L=0$ state beyond $w\sim 4l_B$. 
\bibitem{luhman}  D. R. Luhman, W. Pan, D. C. Tsui, L. N. Pfeiffer, K. W. Baldwin, and K. W. West, Phys. Rev. Lett. {\bf 101}, 266804 (2008); J. Shabani, T. Gokmen, M. Shayegan, Phys. Rev. Lett. {\bf 103} 046805 (2009).
\bibitem{tqc}  C. Nayak, S. H. Simon, A. Stern, M. Freedman, and S. Das Sarma, Rev. Mod. Phys. {\bf 80}, 1083 (2008).
\bibitem{shabani} J. Shabani \emph{et al.}, arXiv:0909.2262.
\bibitem{rh} E. Rezayi and F. D. M. Haldane, Phys. Rev. Lett. {\bf 84}, 4685 (2000).
\bibitem{ms} G. M\"{o}ller and S. H. Simon, Phys. Rev. B {\bf 77}, 075319 (2008).
\bibitem{storni} M. Storni, R. H. Morf and S. Das Sarma, arXiv:0812.2691.
\bibitem{morf_cfsea} R. Morf and N. d'Ambrumenil, Phys. Rev. Lett. {\bf 74}, 5116 (1995).
\bibitem{park}  K. Park, V. Melik-Alaverdian, N. E. Bonesteel, and J. K. Jain, Phys. Rev. B {\bf 58}, R10167 (1998)
\bibitem{rr} E. H. Rezayi and N. Read, Phys. Rev. Lett. {\bf 72}, 900 (1994).
\bibitem{scarola_rezayi_jain} V. W. Scarola, J. K. Jain, and E. H. Rezayi, Phys. Rev. Lett. {\bf 88}, 216804 (2002).

\bibitem{morf_ambrumenil}  R. Morf and N. d'Ambrumenil, Phys. Rev. B {\bf 68}, 113309 (2003).
\bibitem{haldane_entspectrum}  H. Li and F. D. Haldane, Phys. Rev. Lett. {\bf 101}, 010504 (2008).
\bibitem{zhr} O. S. Zozulya, M. Haque, and N. Regnault, Phys. Rev. B {\bf 79}, 045409 (2009);  N. Regnault, B. A. Bernevig, and F. D. Haldane, Phys. Rev. Lett. {\bf 103}, 016801 (2009).

\end{thebibliography}
\end{document}